# Analysis of Ownership and Travel Behavior of Women Who Drive Electric Vehicles: The case of Maryland

*Paper presented at the 6th International Conference on Women's Issues in Transportation, Irvine, CA 2019*

**Amirreza Nickkar, Corresponding Author**
Ph.D. Student
Transportation and Urban Infrastructure Studies, Morgan State University, 1700 E Cold Spring Lane, Baltimore, MD 21251
Email: amirreza.nickkar@morgan.edu

**Hyeon-Shic Shin, Ph.D.**
Assistant Professor
City and Regional Planning, Morgan State University, 1700 E. Cold Spring Lane, Baltimore, MD 21251
Tel: 443-885-1041; Fax: 443-885-8275
Email: hyeonshic.shin@morgan.edu

**Z. Andrew Farkas, Ph.D.**
Director
National Transportation Center, Morgan State University, 1700 E. Cold Spring Lane, Baltimore, MD 21251
Tel: 443-885-3761; Fax: 443-885-8275
Email: andrew.farkas@morgan.edu

**Aknowledgements**: The authors would like to thank the Mid-Atlantic Transportation Sustainability University Transportation Center, led by the University of Virginia, and the U.S. Department of Transportation University Transportation Centers Program for their financial support of this research. The authors also thank the Maryland Motor Vehicle Administration for its cooperation with the survey of electric vehicle owners.

# INTRODUCTION

Electric vehicles (EVs) can bring lots of benefits to the environment such as reduced energy consumption and emissions. Since the U.S. federal government has invested in policies that promote EVs in recent years, it is necessary to have clear insights into the travel behavior and ownership status of the EV owners. Currently in the U.S., men are more than twice as likely as women to own electric cars, although recent studies show that more women are making the switch to drive EVs [1]. Therefore, understanding possible influencing factors on purchasing EVs and travel behavior of EV owners would provide more practical and theoretical insights toward enhancing investment policies in the EV industry. The influence of demographic factors on EV purchasing behavior was extensive in most of the past studies. It has been proved that some socio-demographic characteristics like education [2-4], income [5, 6], gender [7, 8], age [9, 10], and household size [11, 12] may have a significant relationship with the EV preference and adoption. the reader is referred to the excellent review paper by Liao et al. [13] for a comprehensive literature review on this topic. Although many studies tried to address possible influencing factors on purchasing behavior of EV owners and their travel behavior, there is a very little knowledge is available about spatial travel and ownership behavior of EV owners especially in the angle of gender difference.

Several research questions arise from the state's policies to spur EV ownership through subsidizing purchase price and deploying public charging facilities at rail transit stations. Who drives EVs and what are female EV owners' socioeconomic characteristics? What are the primary reasons for females to purchase an EV and how are they related to owners' attitude toward and preferences for purchasing reasons such as environmental concerns, safety, gas prices, vehicle performance, and others? To answer the questions this study aims to investigate the contributing socio-demographic characteristics and factors among female EV owners as well as their travel behavior, commuting trip patterns, and purchasing/leasing ownerships. The objective of the study is to recommend public policies to decision makers to prompt gender equity for EV purchase and use by identifying socio-demographic attributes that influence EV travel patterns and behavior.

# METHODOLOGY

This research surveyed registered Battery Electric Vehicle (BEV) owners in Maryland regarding attitudes toward EV purchasing and travel behavior, environmental considerations, and mode choice for work trips before and after purchase. The method of survey was online and respondents were asked about socioeconomic characteristics, vehicle features, current technology use, travel attributes, and preferences.

An online survey of EV owners of both sexes was conducted from May 28, 2015, to February 19, 2016. In total, 1,257 EV owners in Maryland completed usable surveys. A set of statistical analysis methods was employed to analyze the data [14, 15]. The sample was skewed toward the male population; 75% of the respondents were male, and 25% were female. Table 1 shows a summary of selected socioeconomic characteristics by gender.

**Table 1. EV owners' demographic characteristics by gender**

| Attribute | | Count | | Percentage | |
|---|---|---|---|---|---|
| | | Male | Female | Male | Female |
| Age | 30 years old and younger | 17 | 3 | 85 | 15 |
| | 30 to 49 years old | 236 | 80 | 75 | 25 |
| | 50 to 59 years old | 215 | 66 | 77 | 23 |
| | 60 years old and older | 149 | 50 | 75 | 25 |
| People in Household | One | 49 | 14 | 78 | 22 |
| | Two | 228 | 103 | 69 | 31 |
| | Three or more | 339 | 82 | 81 | 19 |
| Vehicles in Household | One | 49 | 19 | 72 | 28 |
| | Two | 278 | 90 | 76 | 24 |
| | Three or more | 290 | 90 | 76 | 24 |
| Education | College degree, high school diploma, and under | 95 | 17 | 85 | 15 |
| | Bachelor's degree | 177 | 45 | 80 | 20 |
| | Master's degree | 172 | 69 | 71 | 29 |
| | Doctoral or professional degree | 172 | 66 | 72 | 28 |
| Income | Less than $100,000 | 67 | 34 | 66 | 34 |
| | $100,000 – $200,000 | 221 | 83 | 73 | 27 |
| | More than $200,000 | 247 | 56 | 82 | 18 |
| Marital status | Single | 79 | 27 | 75 | 25 |
| | Married or in domestic partnership | 534 | 170 | 76 | 24 |
| Race/Ethnicity | White (non-Hispanic) | 473 | 160 | 75 | 25 |
| | Other | 94 | 31 | 75 | 25 |
| Political affiliation | Democrat | 287 | 128 | 69 | 31 |
| | Republican | 96 | 21 | 82 | 18 |
| | Independent | 153 | 28 | 85 | 15 |
| | Not interested in politics | 70 | 20 | 78 | 22 |

According to the US Census Bureau, geographic areas are classified into urban and rural areas. Urban areas in this classification include cities and towns, where 81% of the US population lives, and the rest (19%) are living in rural areas. However, this classification may not be adequate for analyzing travel behavior and commuting patterns of travelers. In order to conduct a spatial analysis of commuting trips of EV owners, the geographic areas in Maryland were categorized into three levels of "city," "suburban" and "rural" areas. This classification clarifies travel patterns of EV owners. Figure 1 shows geographical distribution of female EV owners in Maryland.

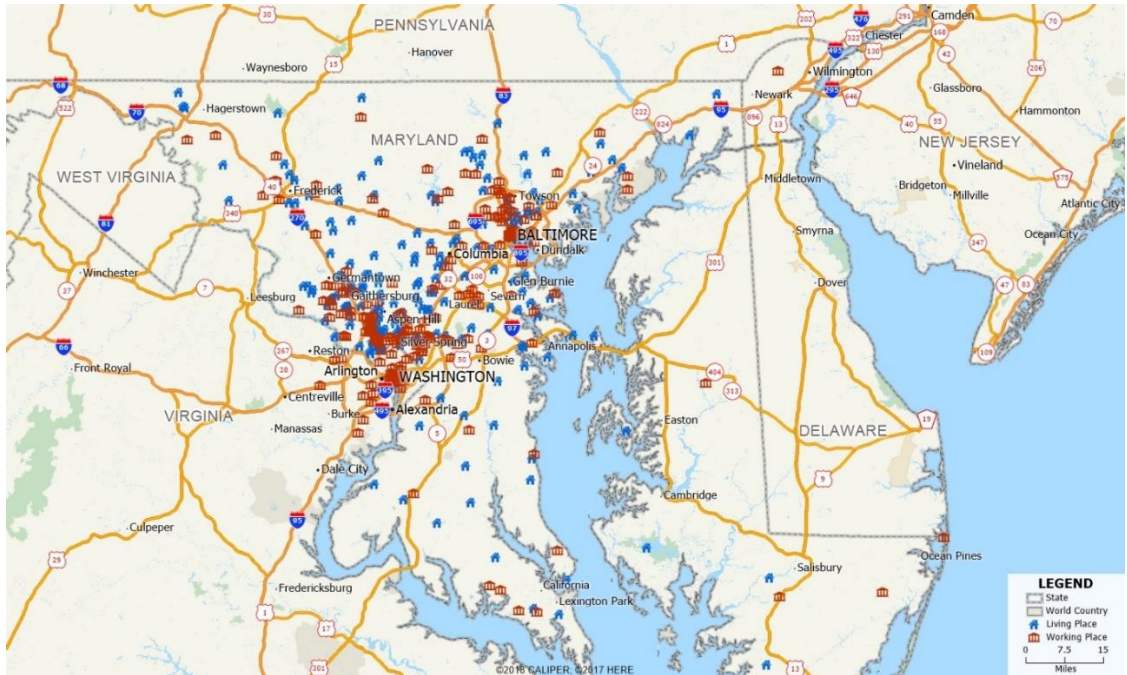
**Figure 1. Geographical distribution of female EV owners in Maryland**

A set of correlation test, analysis of variance (ANOVA), and multinomial logit model (MNL) were constructed to examine the associations between EV owner characteristics and their EV purchasing/leasing and spatial commuting trip behavior. The EV owners in Maryland were asked to select three top reasons that encouraged them to buy or lease an EV.

**FINDINGS**

It is hypothesized that there is a relationship between female EV owners' commuting trip distance and their sociodemographic characteristics; therefore, in this section, a set of one-way analyses of variance was conducted to examine possible relationships between socio-demographic characteristics of EV owners and their driving distance by EV. Values of F and their significance levels in ANOVA are summarized in Table 2.

**TABLE 2 ANOVA on driving distance mileage among socio-demographic variables**

| Variable | F | Sig. |
|---|---|---|
| Age | 0.998 | 0.53 |
| People in Household | 1.313 | 0.205 |
| Vehicles in Household | 1.318 | 0.201 |
| Education | 1.004 | 0.523 |
| Income | 0.885 | 0.68 |
| Marital status | 1.747 | 0.044* |
| Race/Ethnicity | 2.212 | 0.009* |
| Political affiliation | 1.028 | 0.493 |

*$p \leq .05$*

Table 3 shows the main reasons of participants for purchasing EV.

**Table 3. Summary of female participants' reasons for purchasing/leasing an EV**

| Trip status / main reason for purchasing EV | | Environmental Issues | Price and Authority Issues | Efficiency and Performance | Total |
|---|---|---|---|---|---|
| Suburban to Suburban | Count | 41 | 15 | 8 | 64 |
| | % of Total | 20.70% | 7.60% | 4.00% | 32.30% |
| Suburban to City | Count | 32 | 15 | 1 | 48 |
| | % of Total | 16.20% | 7.60% | 0.50% | 24.20% |
| Suburban to Rural | Count | 3 | 0 | 0 | 3 |
| | % of Total | 1.50% | 0.00% | 0.00% | 1.50% |
| City to City | Count | 21 | 9 | 3 | 33 |
| | % of Total | 10.60% | 4.50% | 1.50% | 16.70% |
| City to Suburban | Count | 21 | 3 | 1 | 25 |
| | % of Total | 10.60% | 1.50% | 0.50% | 12.60% |
| City to Rural | Count | 2 | 0 | 0 | 2 |
| | % of Total | 1.00% | 0.00% | 0.00% | 1.00% |
| Rural to City | Count | 4 | 2 | 1 | 7 |
| | % of Total | 2.00% | 1.00% | 0.50% | 3.50% |
| Rural to Suburban | Count | 5 | 5 | 1 | 11 |
| | % of Total | 2.50% | 2.50% | 0.50% | 5.60% |
| Rural to Rural | Count | 3 | 0 | 2 | 5 |
| | % of Total | 1.50% | 0.00% | 1.00% | 2.50% |
| Total | Count | 132 | 49 | 17 | 198 |
| | % of Total | 66.70% | 24.70% | 8.60% | 100.00% |

Table 4 presents Chi-square and statistical significance of variables derived from likelihood ratio tests.

**Table 4. Results of Likelihood Ratio Tests for EV purchasing reasons logit model**

| Likelihood Ratio Tests | | | |
|---|---|---|---|
| | Model Fitting Criteria | Likelihood Ratio Tests | |
| Effect | -2 Log Likelihood of Reduced Model | Chi-Square | Sig. |
| Intercept | 134.444[a] | 0.000 | |
| Age*Education | 175.957 | 41.513 | 0.000* |
| Age*Vehicle | 155.355 | 20.911 | 0.022* |
| Political affiliation | 161.614[b] | 27.170 | 0.000* |
| Education | 153.908 | 12.982 | 0.043* |
| The chi-square statistic is the difference in -2 log-likelihoods between the final model and a reduced model. The reduced model is formed by omitting an effect from the final model. The null hypothesis is that all parameters of that effect are 0. | | | |
| a. This reduced model is equivalent to the final model because omitting the effect does not increase the degrees of freedom. | | | |
| b. Unexpected singularities in the Hessian matrix are encountered. This indicates that either some predictor variables should be excluded or some categories should be merged. | | | |

* $p \leq .05$

## CONCLUSIONS

The results of this study showed that first, socio-demographic factors, including education and income, played a significant role in preferences attributes of participants for purchasing/leasing an EV and in the commuting travel behavior and pattern and of EV drivers. Second, environmental issues are the main reason for purchasing/leasing EVs, but the EV owners who had longer commutes were more concerned about the price and status of the EV owner and efficiency and performance than were those with shorter commutes. The results of logit models showed that females with higher age, income, and education levels were more concerned about the environmental issues than younger ones. There is a big gender gap in EV ownership in Maryland. Several reasons for male dominance could be speculated. It is possible that most households registered their EVs under male householders, and males were likely to be primary EV drivers.